# A Re-analysis of Repeatability and Reproducibility in the Ames-USDOE-FBI Study

**A Re-analysis of Repeatability and Reproducibility in the Ames-USDOE-FBI Study**

**Alan H. Dorfman [a,*], Richard Valliant [b]**

[a] *National Center for Health Statistics (retired), Bethesda, MD 20814, USA*

[b] *Research Professor Emeritus, Universities of Michigan & Maryland, Ann Arbor MI 48104*

\* Corresponding author

E-mail addresses: dorfmans@erols.com, valliant@umich.edu

*Abstract*

Forensic firearms identification, the determination by a trained firearms examiner as to whether or not bullets or cartridges came from a common weapon, has long been a mainstay in the criminal courts. Reliability of forensic firearms identification has been challenged in the general scientific community, and, in response, several studies have been carried out aimed at showing that firearms examination is accurate, that is, has low error rates. Less studied has been the question of consistency, of. whether two examinations of the same bullets or cartridge cases come to the same conclusion, carried out by an examiner on separate occasions—intrarater reliability or *repeatability*—or by two examiners—interrater reliability or *reproducibility*.

One important study, described in a 2020 Report by the Ames Laboratory-USDOE to the Federal Bureau of Investigation, went beyond considerations of accuracy to investigate firearms examination repeatability and reproducibility. The Report's conclusions were paradoxical. The observed agreement of examiners with themselves or with other examiners appears mediocre. However, the study concluded repeatability and reproducibility are satisfactory, on grounds that the observed agreement exceeds a quantity called the *expected agreement*. We find that appropriately employing expected agreement as it was intended does *not* suggest satisfactory repeatability and reproducibility, but the opposite.

**1. Introduction**

This paper reviews some results on forensic firearms comparisons appearing in *Report: Validation Study of the Accuracy, Repeatability, and Reproducibility of Firearms Comparisons October 7, 2020* (Bajic et al. 2020)[1], conveying results of a study by the Ames Laboratory at Iowa State University under the auspices of the U.S. Department of Energy at the behest of the U.S. Federal Bureau of Investigation (henceforth "Report" or "Ames-FBI Study").  We focus here on just one aspect of the study, that bearing on *repeatability* and *reproducibility*—pages 37 – 52 of the Report.

Speaking generally, the *reliability* of a diagnostic procedure—in general, a procedure meant to translate an observation or set of observations into one category in a set of categories—is determined by looking at its *accuracy*, how often the true category is selected, and also by its consistency as manifested in its *repeatability*, how often separate observations by the same instrument or person on the same objects arrive at the same classification, and its *reproducibility*, how often different instruments or persons observing the same objects arrive at the same classification.  We note that repeatability and reproducibility are weaker conditions than accuracy—a finding of high repeatability or reproducibility means only that measurements give consistent results.  They could be consistently wrong.  On the other hand, inadequate repeatability or reproducibility implies that the procedure cannot always be on target.

What is *adequate* repeatability or reproducibility?  The answer must depend on the circumstances (McHugh 2012)   If the diagnosis or judgment bears on a minor matter, for example, whether the upcoming winter's local snowfall will be meager, average, or heavy, we will not be troubled if different assessments availing of the same information don't have tight agreement (they probably won't). In more serious matters, for example, if a person's quality of life, or life itself, is at stake, as in some medical and forensic situations, anything short of a very high degree of consistency should raise doubts about the procedure.

---

[1] This Report is unfortunately not available in the public domain as of this writing. It was released to the public in early 2021 and then in the course of the year withdrawn.  Before being withdrawn,  it circulated widely, and continues to be put into evidence in court cases throughout the U.S., for example, *State v. Gregory Jones (2021),  US v. Mario Felix (2022)* . We discuss it therefore for its importance, and the particular light its sheds on the question of repeatability and reproducibility.

In its most prominent usage, the process of forensic firearms examination is tasked with the comparison of bullets (or cartridges) with each other to determine whether or not they derive from the same gun or not. For example, an examiner may be asked to assess whether bullets recovered from a crime scene were fired by a gun owned by a suspect. An examiner then testifies in court on his/her conclusions. A jury will weigh that testimony, along with any other evidence presented in a trial to determine a suspect's guilt or innocence. The firearms examiner compares the markings on pairs of bullets or cartridges under a specialized microscope and, following guidelines of the Association of Firearms and Toolmarks Examiners (AFTE Criteria for Identification Committee 1992), arrives at one of six conclusions, as shown in Figure 1.

The AFTE Range of Conclusions was developed by the Criteria for Identification Committee and adopted by the Association membership at its annual business meeting in April 1992 (and published in the AFTE Journal Volume 24, Number 3).

1. Identification
   Agreement of a combination of individual characteristics and all discernible class characteristics where the extent of agreement exceeds that which can occur in the comparison of toolmarks made by different tools and is consistent with the agreement demonstrated by toolmarks known to have been produced by the same tool.
2. Inconclusive
   a. Some agreement of individual characteristics and all discernible class characteristics, but insufficient for an identification.
   b. Agreement of all discernible class characteristics without agreement or disagreement of individual characteristics due to an absence, insufficiency, or lack of reproducibility.
   c. Agreement of all discernible class characteristics and disagreement of individual characteristics, but insufficient for an elimination.
3. Elimination
   Significant disagreement of discernible class characteristics and/or individual characteristics.
4. Unsuitable
   Unsuitable for examination.

Figure 1.  The AFTE Range of Conclusions

Different forensic laboratories and examiners adopt the AFTE criteria differently, for example, ignoring the distinctions among the Inconclusives, or restricting Eliminations to those based on class characteristics (Baldwin et al 2016., p. 7).

Testimony based on firearms examination have been accepted in courts for over a century.  However, in recent times the studies purporting to establish reliability of the conclusions has been called into question by respected scientific bodies, most notably a 2016 Report by the President's Council of Advisers on Science and Technology (2016)--PCAST, which emphasized accuracy, repeatability and reproducibility as the ingredients of reliability (PCAST, Box 2, p. 47).

Almost all firearms studies have focused purely on accuracy.   We are aware of only two other studies that have considered reproducibility.   Law and Morris (2021) (using plastic casts of cartridge cases) and Kerstolt et al.(2010) (bullets).  In both cases reproducibility was modest:  "Variability in examiner conclusions when evaluating the same evidence was observed" (Law and Morris, p.1);  Kerstolt et al. found pairs of examiners agreed in 84% of their comparisons of same bullets (p. 141).

The Ames-DOE-FBI Report is groundbreaking among firearms studies in employing a study design complex enough to allow a thoroughgoing assessment of repeatability and reproducibility. Although the Report also addresses accuracy (and other matters), we limit our focus in this review to the methods, results and line of argument concerning repeatability and reproducibility.  For the sake of compactness, we will typically go into detail only regarding results for bullets;  cartridges will differ slightly in the numbers, but the general analysis and the overall conclusions are the same.

We describe the study's methodology, results and conclusions, with special attention to the concept of *expected agreement*.   The authors of the Report find that the agreement actually observed generally exceeds a quantity called *expected agreement*, and from that conclude that repeatability and reproducibility are good.   We have no reason to doubt the finding, but are unpersuaded as to the conclusion, which, we think, misappropriates expected agreement.

**2. Study Design**

The Ames-FBI Study was the second large experiment on firearms comparison to be carried out by the Ames Laboratory at Iowa State University.  The first, "Ames-I" (Baldwin et al 2016) had involved only cartridges.  In contrast to previous firearms studies, Ames-I had received good marks from PCAST for its open, independent design—"open" meaning that comparison sets included cartridges for which there were no other cartridges from the same gun, "independent" meaning that comparisons were made within a succession of separate small sets.  Previous studies had tended to be large collections of bullets or cartridges, and examiners were tasked with matching up those bullets or cartridges which came from the same gun; each item had at least one match. This had the effect of allowing for a type of deduction and elimination that wouldn't be typically available in casework.  Set to set comparisons also complicate the calculation of error rates;  compare (Hofmann et al 2020, p. 337, and Appendix C).

The Ames-FBI Study was an expanded, more ambitious version of Ames-I study.  It included bullets as well as cartridges, more than one brand of gun, and, most distinctively, it allowed for the masked issuing of the same bullets/cartridges more than once to the same or a different examiner.  The intention was to be able to assess the extent to which, on repeated examination, an examiner came to the same conclusion (intra-rater reliability or *repeatability*), and likewise, the extent to which different examiners agreed on the same items (inter-rater reliability or *reproducibility*).

The Ames Laboratory broadcast an invitation to firearms examiners, seeking volunteers for the study. There were some limitations: the examiners had to belong to AFTE (Association of Firearms and Toolmarks Examiners) or work for an accredited forensic laboratory.  Because they were involved in the design and material development of the study, members of the FBI were excluded, to avoid bias. The letter of invitation (Report, Appendix A) apprised examiners of the length and difficulty of the study, and the goal of measuring accuracy, repeatability and reproducibility.  It also allowed that dropping out of the study at any point was an option.

Those who stayed in for the duration of the study received six (6) mailings over two years, each containing a packet of 15 sets of bullets and a packet of 15 sets of cartridges.  Each set of bullets/cartridges consisted of two "knowns" from one gun and one "unknown" from either that gun ("Matching") or a different gun of the same brand and model ("Nonmatching").  The ratio of matches to non-matches varied across packets, with 3 - 7 matches and 8 - 12 nonmatches among the 15 sets. Packets also varied in the extent the two different brands of guns used were the source of bullets/cartridges. The Report suggests the guns and ammunition were chosen to make comparisons difficult compared to the general spectrum of difficulty of comparisons an examiner would face in practice.

An examiner was advised to choose between 6 conclusions, in accord with the full AFTE classification that breaks inconclusives into sub-categories: Identification, (judgment that the unknown came from the same source as the knowns), Elimination (it didn't), three variants of Inconclusive-- leaning to inclusion (A), leaning to elimination (C), and non-leaning (B)--and Unsuitable, in which the unknown or both knowns were lacking in sufficient markings to allow for comparison. (see Figure 1 above) After completing his or her comparisons, the examiner would return the bullets and cartridges along with his or her conclusions to the Ames laboratory, allowing for the re-issue of the same packets at a later time to the same or another examiner.

The mailings were grouped into "rounds"—Round 1 to assess accuracy, Round 2 for repeatability, Round 3 for reproducibility. A mailing could be regarded as part of more than one round, but any two mailings in Round 2, the round used to assess repeatability, were separated by at least one intervening mailing, to disguise the re-issue of items to an examiner.

Initially, 250 examiners responded to the letter of invitation and received packets. Of these, 173 followed through and reported conclusions on the packets of the first mailing (an attrition rate of 31%, itself raising concerns, but beyond the scope of this paper). 105 examiners (42%) remained in the study long enough to provide grist for estimates of repeatability and reproducibility. The examiners who participated ranged widely in their years of experience, from less than a year to fifty years, with a median of 9 years.

## 3. Results

*3.1 Repeatability*

Data on paired classifications by the same examiner (Repeatability) for Bullets (Report, Table IX, p. 38), here repeated as "Table 9", gives the raw data, collected over examiners, for both Matching Sets (where the unknown is from same gun as the two knowns) and Nonmatching (where the unknown is from a different gun than shot the two knowns). This Table was based on the subset of examiners who stayed in for Round 2 (mailings 3-6). The number of conclusions that were the same both rounds (sum of the bolded elements on diagonal of the table) divided by the total number of comparisons (sum of all the numbers in the Table), yields "...a rough measure of reliability agreement,..." (Report, p. 39). For matching sets, this is 758/960 = 79.0% and, for nonmatching, 1201/1855 = 64.7%. In other words, for bullets, there was self-disagreement 21.0% and 35.3 % of the time, for matching and nonmatching respectively. Results for cartridges are similar.

**Table 9. Paired Classifications by the Same Examiner (Repeatability) for Bullets.**

### Matching sets -Bullets

|                | Identification | Inconclusive A | Inconclusive B | Inconclusive C | Elimination | Unsuitable | Total |
|----------------|---------------:|---------------:|---------------:|---------------:|------------:|-----------:|------:|
| Identification | **665**        | 27             | 26             | 14             | 8           | 2          | 742   |
| Inconclusive A | 31             | **28**         | 12             | 6              | 2           | 0          | 79    |
| Inconclusive B | 13             | 14             | **45**         | 5              | 2           | 2          | 81    |
| Inconclusive C | 2              | 3              | 3              | **5**          | 3           | 0          | 16    |
| Elimination    | 8              | 7              | 3              | 2              | **13**      | 0          | 33    |
| Unsuitable     | 1              | 3              | 3              | 0              | 0           | **2**      | 9     |
| Total          | 720            | 82             | 92             | 32             | 28          | **6**      | 960   |

### Nonmatching sets -Bullets

|                | Identification | Inconclusive A | Inconclusive B | Inconclusive C | Elimination | Unsuitable | Total |
|----------------|---------------:|---------------:|---------------:|---------------:|------------:|-----------:|------:|
| Identification | **2**          | 3              | 6              | 2              | 6           | 0          | 19    |
| Inconclusive A | 0              | **52**         | 37             | 42             | 27          | 0          | 158   |
| Inconclusive B | 5              | 31             | **341**        | 98             | 45          | 7          | 527   |
| Inconclusive C | 1              | 32             | 109            | **284**        | 53          | 1          | 480   |
| Elimination    | 1              | 20             | 35             | 66             | **514**     | 4          | 640   |
| Unsuitable     | 0              | 0              | 13             | 6              | 4           | **8**      | 31    |
| Total          | 9              | 138            | 541            | 498            | 649         | **20**     | 1855  |

Note: From Report, Table IX, p. 38

This level of disagreement seems high. One can note some anomalies of detail as well: among the Matches there are three instances where on one of the two rounds an examiner or examiners judged the bullets to be unsuitable—not enough information to even make a comparison—and on the other saw an identification (upper right and lower left corners of the table.) There were 16 instances (8 Identification/Elimination + 8 Elimination/Identification) of an elimination being called an identification on the alternate round. Among Nonmatches, there were 35 instances of unsuitables on one round falling into another category on the other; there were 7 instances of eliminations being classified as identifications on the alternate round; there were 47 instances of eliminations being judged as leaning towards identification (the Inconclusive-A's.)

The authors note that some of the disagreement could be due to relatively inconsequential shifts between the three varieties of Inconclusive, and they investigate what happens if (a) the Inconclusives are lumped together, and (b) if the A-type is coalesced with Identification and the C-type with Elimination. The results are given in **Table XII** (Report, p. 42), repeated in Table 12 below, Under (a)

**Table 12: Proportion of Paired Classifications in Agreement and Disagreement by the Same Examiner (Repeatability) when (a) the three Inconclusive Categories are Pooled, (b) ID and Inconclusive-A are Pooled and Elimination and Inconclusive-C are Pooled -Bullets.**

|  | Proportion of paired agreements | Proportion of paired disagreements |
|---|---|---|
| **Matching Sets** | | |
| **(a) Pooling of Inconclusives** | 83.4% | 16.6% |
| **(b) Pooling of ID with Inconclusive-A and of Elimination with Inconclusive-C** | 85.5% | 14.5% |
| **Nonmatching Sets** | | |
| **(a) Pooling of Inconclusives** | 83.6% | 16.4% |
| **(b) Pooling of ID with Inconclusive-A and of Elimination with Inconclusive-C** | 71.3% | 28.7% |

Note: Based on Report, Table XII, p. 42

the rate of disagreement is reduced to 16.6% for Matching and 16.4% for Nonmatching; for (b), to 14.5% for Matching and 28.7% for Nonmatching. This means that on average the examiners came to a different conclusion on a second comparison of a set of bullets, at least once in 7 comparisons, however one merges the inconclusives. Results for cartridges are similar.

The amount of disagreement under any scenario seems substantial, and certainly gives us the impression of not very strong repeatability. The percentages for cartridges are similar.

*3.2 Reproducibility*

Not surprisingly, reproducibility—the tendency of *different* examiners to come to the same conclusion on a given set of bullets—is lower than that for repeatability.

Table XIV: paired Classifications by Different examiners (Reproducibility) for Bullets (Report, p. 46, not repeated here), gives the raw data across examiners. Making the same sort of calculations—the sum of the numbers on the diagonal divided by the sum of numbers in the table as a whole—as were carried out for repeatability, reveals a 32.2% disagreement between examiners for Matching sets and an extraordinary 69.1% for Nonmatching. The results for cartridges were similar, 36.4% and 59.7%. (Report, Table XVI, p. 47).

Pooling inconclusives as was done for repeatability, the Report finds, for bullets, the disagreement when inconclusives are merged together—case (a)—to be 27.6% for Matching and 45.4% for Nonmatching. In case (b), where Inconclusive-A is grouped with Identification and Inconclusive-C with Elimination, the disagreement is 22.6% and 51.0% for Matching and Nonmatching respectively. For cartridges. the corresponding percent disagreement is (a) 29.7% for Matching and 45.1% for Nonmatching and (b) 23.6% for Matching and 40.5% for Nonmatching. (Report, Table XVII, p. 49)

These are rather large percentages of disagreement.

**4. Report's Appraisal of Results on Repeatability and Reproducibility**

Overall. the results of the study suggest that, at least for the examiners represented in the study and for the guns and ammunition examined, repeatability and reproducibility are at best mediocre. The level of repeatability and reproducibility as measured by the between rounds consistency of conclusions would not appear to support the reliability of firearms examination.

The authors of the Report do *not* draw the above conclusion. Turning to an analysis based on looking at participating examiners one by one and comparing the observed level of agreement to a measure called the *expected* agreement, they find that, as a rule, observed agreement exceeds that of expected agreement and they read this as indicating that repeatability and reproducibility *are* satisfactory.

The argument begins on page 39 of the Report:

> To separate the examiner-specific effects of repeatability, separate 6x6 tables (as in Tables IX and X) were constructed for each of the 105 examiners who performed repeated examinations of the same sets of bullets and cartridge cases in Rounds 1 and 2. Two statistics were computed from each of these individualized tables:
>
> 1. The proportion of *Observed Agreement*, i.e. the proportion of counts in the table that fall in the shaded cells along the top-left-to-bottom-right diagonal, when both examinations resulted in the same conclusion. These are the count totals re-expressed as the proportion of paired agreements as in Table XI, but for individual examiners.
>
> 2. The proportion of *Expected Agreement*, computed as the sum of corresponding marginal proportions, i.e. the proportion of Identification determinations in the first round times the proportion of Identification determinations in the second round, plus the proportion of Inconclusive-A determinations in the first round times the proportion of Inconclusive-A determinations in the second stage, etc.

The second of these computed statistics requires a bit of explanation. If each examination were entirely *independent* of the others – i.e. if the probability of concluding Identification in the second examination were exactly the same regardless of how the set was evaluated in the first examination – the proportion of Expected Agreement would be an estimate of the number of sets on which the examiner should be expected to agree with him/herself. Hence, **if the proportion of Observed Agreement regularly exceeds the proportion of Expected Agreement, this is an indication of examiner repeatability**. (Report, pp 39-40, emphasis supplied)

The Report goes on to display the contrast between observed and expected in several figures—Figures 11-13 for repeatability, and Figures 13-16 for reproducibility. As an example, the figures for bullets with no pooling of inconclusives is copied below (from **Figure 11**, p. 41). Each point in any of the graphs represents an examiner, and whenever the point is above the drawn 45 degree line, it means that the examiner's observed level of agreement exceeds his/her expected agreement. It can be seen that the vast majority of the examiners lie above the line in all the figures, and this is taken as affirming repeatability and reproducibility.

The authors include without comment "box and whiskers" plots Tukey (1977) of examiners' expected agreement above, and observed agreement to the right side, of the main figure. For example, in the Figure 11b for Nonmatching, in the box-plot on right the heavy horizontal line indicates that the median of the observed agreements was about 0.6, meaning that half of the examiners had observed agreement less than or equal to 60%. The placement of the base of the "box" indicates that 25% of the observed agreements lay below about 0.5—a quarter of the examiners disagreed with themselves half the time.

Table XIII, page 45, gives averages of expected and observed over the several examiners for Repeatability, and it is seen that in every case—matches or nonmatches, various treatments of inconclusives, whether bullets or cartridges—average of the observed agreement exceeds the average of the expected agreement; for example, for bullets:

Table 13. Proportions of Agreement: Bullet Sets

| Scoring | Matches | | Nonmatches | |
| --- | --- | --- | --- | --- |
| | Observed Agreement | Expected Agreement | Observed Agreement | Expected Agreement |
| ID Inc-A Inc-B Inc-C Elim | 77.70% | 63.70% | 63.40% | 55.80% |
| ID (Inc-A & Inc-B & Inc-C) Elim | 82.00% | 66.40% | 82.50% | 77.00% |
| (ID & Inc-A) Inc-B (Inc-C & Elim) | 85.00% | 75.70% | 70.40% | 63.90% |

Note: From Report, Table XIII, p. 45

The authors also do some hypothesis testing:

> A nonparametric sign test of the null hypothesis that observed frequency minus expected frequency has a median of zero, versus the alternative that observed frequency exceeds expected frequency more than half the time, is rejected with a p-value of 0.0001 or less for all data sets represented in Figures 11-13, i.e. bullets and cartridge cases, matching and nonmatching sets, and all three scoring schemes),...(Report, p. 44)

Thus, observed agreement exceeds expected agreement from every point of view.

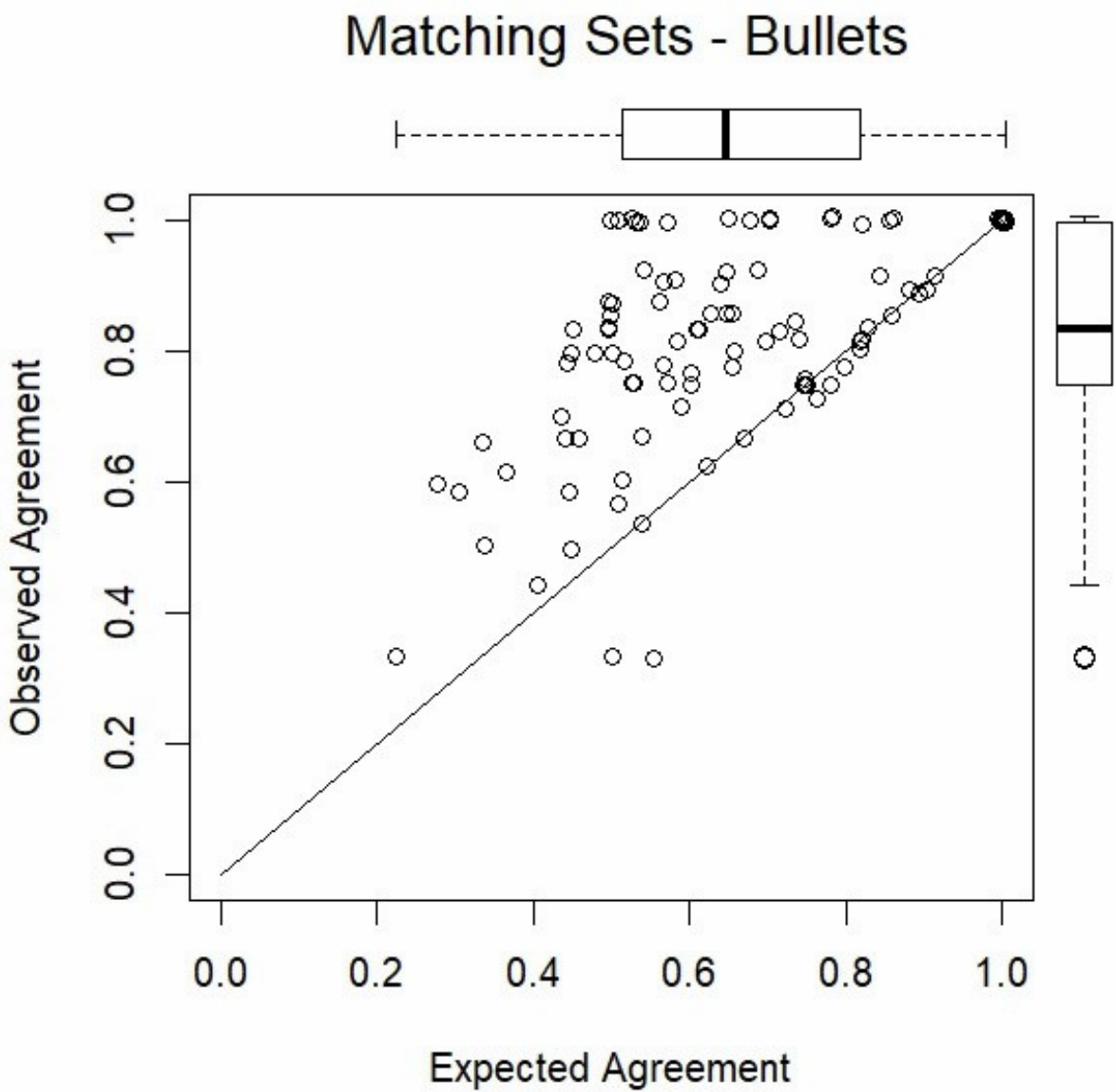

Figure 11 a Repeatability. No Pooling of Inconclusives

(Report, p. 41)

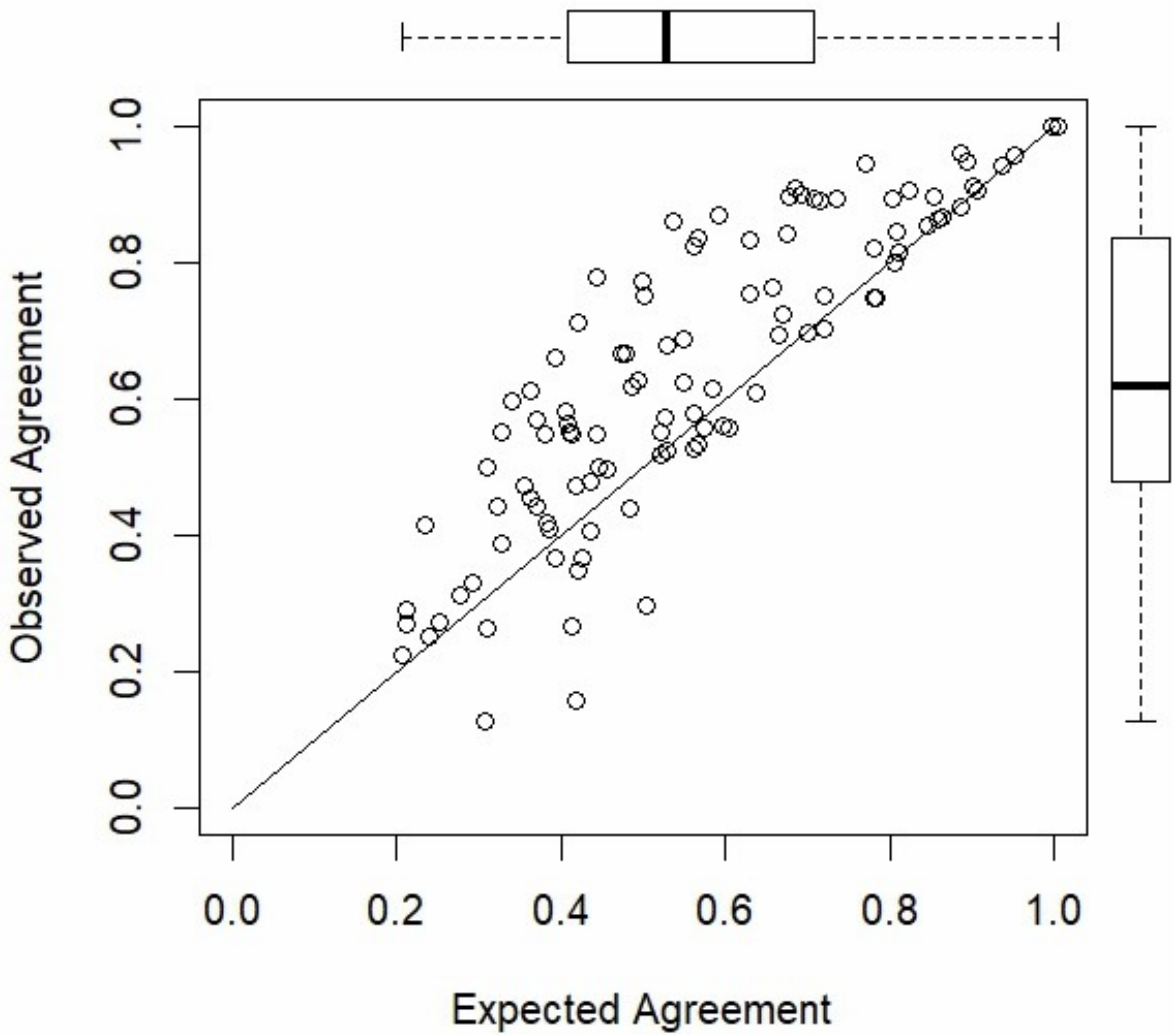

Figure 11 b Repeatability. No Pooling of Inconclusives

(Report, p. 41)

In summing up, the authors state:

> When considering examiner repeatability the plots of Figures 11-13 show that **examiners score high in repeatability**, i.e. their observed performance generally exceeds the statistically expected agreement by a fairly wide margin. This is true whether all three inconclusive category are regarded separately, are pooled as a single category, or Identification and Inconclusive-A results are pooled and Elimination and Inconclusive-C are pooled. (Report, p. 73, emphasis supplied)

The authors go through a similar development for Reproducibility, with similar conclusions (Report, Figures 14 -16 and Table XVIII, pp. 48 - 52). Once again, Observed Agreement exceeds Expected Agreement. Testing results again generally reject the hypotheses that median of observed agreement minus expected agreement is zero, except possibly in one case. And they arrive at similar conclusions:

> ...Figure 14 shows that in general the determinations made by different examiners are reproducible, i.e. the observed agreement falls above the expected agreement line....the general trend toward better observed agreement than expected agreement documents commonality in how the examination process is performed within the profession. (Report, p. 74)

What is this *expected agreement* that it should play such an influential role that it can reverse our initial impressions? A little bit of clarification and historical context will be helpful. Further analysis leads to conclusions different from those in the Report, and, in fact, reinforces the initial impression that repeatability (and *a fortiori* reproducibility) is *not* satisfactory.

## 5. Expected Agreement

*5.1 Definition*

As noted above, the authors begin their discussion of expected agreement: "The second of these computed statistics [proportion of expected agreement] requires a bit of explanation." (Report, p. 39) They go on to mention assumptions of independence and other conditions and they illustrate the concept with an example in Appendix H, in the very last pages of the Report.

It will be helpful to restate the idea of expected agreement in a way slightly different from theirs (although mathematically equivalent), provide some historical context to it, in particular its original, motivating use as a component of an index of agreement called *Cohen's Kappa*, and then re-assess its implications for the study itself.

Consider again Table 9 for Non-Matches, given above. For purposes of illustration, it will be convenient to imagine it as representing the repeatability table for one "mega-examiner" doing a great many comparisons. Dividing each entry by the sum of all the entries in the table we arrive at a table of proportions:

**Table 9.1. Agreement Table of Observed Percentages (%) for Bullets Non-Matching**

|               | Identification | Inconclusive A | Inconclusive B | Inconclusive C | Elimination | Unsuitable |
|---------------|----------------|----------------|----------------|----------------|-------------|------------|
| Identification | **0.11**       | 0.16           | 0.32           | 0.11           | 0.32        | 0          |
| Inconclusive A | 0              | **2.8**        | 1.99           | 2.26           | 1.46        | 0          |
| Inconclusive B | 0.27           | 1.67           | **18.38**      | 5.28           | 2.43        | 0.38       |
| Inconclusive C | 0.05           | 1.73           | 5.88           | **15.31**      | 2.86        | 0.05       |
| Elimination    | 0.05           | 1.08           | 1.89           | 3.56           | **27.71**   | 0.22       |
| Unsuitable     | 0              | 0              | 0.7            | 0.32           | 0.22        | **0.43**   |

This table contains all the relevant information in the original Table 9 (which we can always reconstruct from it provided we know the sum of entries in the original table). If we add the percentages on the diagonal (the bolded numbers) we arrive at the percentage of times the "mega-examiner" agrees with himself—the observed agreement 64.74%, the same as calculated earlier using the raw (frequency) data.

From this table, we can also derive *expected agreement*. It may clarify to do this in steps, making assumptions clear. If we add the percents in a given *row*, we get the percent of times the corresponding classification was arrived at in Round 1. For example the Round 1 percent of *identifications* was 0.11 + 0.16 + 0.32 + 0.11 + 0.32 + 0.00 = 1.02 percent.

In general, the row sums—the so-called marginals—are (in percents)

| idn | incon-A | incon-B | Incon-C | elim | unsuit |
|---|---|---|---|---|---|
| 1.02 | 8.51 | 28.41 | 25.88 | 34.51 | 1.67 |

and if we think of the examiner in Round 1 as randomly allotting the different source bullets to a category, then these row sums can be thought of as (estimates of) the probabilities with which he/she does so. Likewise, adding the columns, we find the marginal probability of second round allotments to be

| idn | incon-A | incon-B | Incon-C | elim | unsuit |
|---|---|---|---|---|---|
| 0.48 | 7.44 | 29.16 | 26.84 | 35.00 | 1.08 |

In general, two events *A* and *B* are *statistically independent*, if the probability of their both occurring equals the *product* of their individually occurring (Feller, 1977, Chap 5, Section 3, pp. 125 seq). For example, the probability that a fair coin will be heads at the same time a fair die turns up '3' (or any other of the six possibilities) is 1/2 times 1/6 = 1/12 = 8.33%. Likewise, on the assumption that the outcomes of the two rounds are in accord with the above marginal probabilities and statistically independent, we calculate the probability of, say, the joint occurrence of Inconclusive-C in the first round and Elimination in the second round as (25.88/100)(35/100) = 9.05 %.

In this way we can construct a matrix of joint probabilities first and second round, on the assumption the two rounds are carried out independently and in accord with the respective marginal probabilities in each round:

**Table 9.2. Agreement Table of Expected Proportions (%) – Bullets Non-Matching**

|  | Identification | Inconclusive A | Inconclusive B | Inconclusive C | Elimination | Unsuitable |
|---|---|---|---|---|---|---|
| Identification | **0** | 0.08 | 0.3 | 0.27 | 0.36 | 0.01 |
| Inconclusive A | 0.04 | **0.63** | 2.48 | 2.29 | 2.98 | 0.09 |
| Inconclusive B | 0.14 | 2.11 | **8.29** | 7.63 | 9.94 | 0.31 |
| Inconclusive C | 0.13 | 1.93 | 7.55 | **6.95** | 9.05 | 0.28 |
| Elimination | 0.17 | 2.57 | 10.06 | 9.26 | **12.07** | 0.37 |
| Unsuitable | 0.01 | 0.12 | 0.49 | 0.45 | 0.58 | **0.02** |

Then, just as the observed agreement was the sum of the diagonal elements in the original table, so we regard the sum of the diagonal elements of the joint probability table (0+ 0.63+ etc.= 27.96%) as the *expected* agreement—what the overall agreement would be if outcomes in rounds one and two were statistically independent and distributed according to the marginal probabilities.

*5.2 Cohen's Kappa*

Expected agreement was introduced in a landmark paper by Jacob Cohen (1960), who was interested in deriving a measure of agreement between two agents forming categorical judgments on the same material. After having displayed a table framed as in Table 9.1 comparing results of two observers, he seeks a measure of how much the two agree. "The most primitive approach has been to simply count up the proportion of cases in which the judges agreed..." (Cohen, p. 38] –just the sum of the diagonal percentages, what we have been referring to as the observed agreement.

Cohen is not content with this as a measure. He goes on: "It takes relatively little in the way of sophistication to appreciate the inadequacy of this solution.....A certain amount of agreement is to be expected by chance, which is really determined by finding the joint probabilities of the marginals. (p. 38)" This is important. He is implying the observed agreement *overstates* the real degree to which the judges agree: some of the agreement is happenstance. What Cohen seeks is the excess over chance agreement, and to measure that he suggests what has come to be called *Cohen's Kappa* = (observed % agreement − expected % agreement)/ (100% − expected % agreement ). In symbols,

$$\kappa = \frac{P_o - P_e}{1 - P_e}, \qquad (1)$$

where $P_0$ is the observed agreement and $P_e$ is the expected agreement (Cohen, p. 40).

Thus, *kappa* is a way of discounting chance in measuring agreement. It will always be less than the observed agreement, except for being 100% when observed agreement itself is 100%. It will be *zero* when observed agreement equals expected agreement   It can be negative.

In the Figures of the Report, such as the Figures 11 depicted above, those points lying on the drawn (45 degree) line will have *kappa* = 0. All points lying above that line will have *kappa* greater than 0. There are a few lying below the line where observed agreement is actually less than expected agreement, where, so to speak, the examiner's consistency doesn't match what he would have gotten by good guessing.

What is a satisfactory *kappa*? This has been debated with various standards suggested over the years,

for example McHugh (2012) who, in the context of comparisons affecting health, suggests:

**Interpretation of Cohen's kappa (McHugh 2012)**

| Value of Kappa | Level of Agreement |
|---|---|
| 0–.20 | None |
| .21–.39 | Minimal |
| .40–.59 | Weak |
| .60–.79 | Moderate |
| .80–.90 | Strong |
| Above .90 | Almost Perfect |

In particular, the mere fact that kappa exceeds 0—that observed agreement exceeds expected—does not seem particularly informative. Indeed the originator of kappa didn't think so: "It needs pointing out that is is generally as of little value to test kappa for significance as it is for any other reliability coefficient—to know merely that kappa is beyond chance [i.e. greater than *zero*] is trivial since one usually expects much more than this in the way of reliability...." (Cohen, p.44). Nor did A. Agresti, the author of the classic text on dealing with categorical data: "It is rarely plausible that agreement is no better than chance." (Agresti 2013, p. 438) In the context of the Ames-OES-FBI Report, this is saying that we should nowise be surprised at the Results in Tables 13 and 18, nor that, in Figures 11 - 16, the great bulk of examiners fall above the. *kappa* = 0 line.

What *kappa* might be a good cutoff? What values of *kappa* suggest repeatability or reproducibility are satisfactory?

Appendix A illustrates expected agreement and kappa in a simplified, paradigmatic, situation, where conclusions are based on a mix of clear-sightedness and well-founded guessing—the "shrewd guessing model". Expected agreement turns out then to be the equivalent of the observed agreement of a *blind* observer, astutely guessing. *Kappa* represents the degree of clear sightedness, and 1 – *kappa* measures the degree of guessing, as proven in *Appendix B*. Interpreting *kappa* this way can give us a handle on how satisfactory *kappa* is in particular circumstances. In something as critical to the justice system as firearms examination, we probably want to be very demanding; anything less than an observed agreement of 90%, or a *kappa* of 80% seems grounds for questioning the reliability of firearms comparisons, certainly cannot be taken as supporting that reliability.

Data from the study at the examiner level has not yet been made available, so we are unable to calculate the *kappa* indexes for each of the 105 examiners for the different scenarios the study presents. We can, however, get some idea where they lie by a simple device. From equation (1), we can, for fixed *kappa*, write the observed agreement *Po* as a linear function of the expected agreement *Pe*:

$$P_o = (1 - \kappa)P_e + \kappa \qquad (2)$$

That is, *kappa* is the intercept and 1- *kappa* is the slope of the line on which lie the points where the expected and observed coordinates yield that value of *kappa*. Points (corresponding to examiners) lying above the line will have greater kappa, those below, less. The authors have graphed the (as we have seen, not too meaningful) line corresponding to *kappa* = 0.

It is of interest to add the lines corresponding to *kappa* = 80% to the figures depicting Observed Agreement versus Expected Agreement, as in Figures 12a and 12b below, taken from the Report's Figure 12 (p. 43), where the inconclusives are pooled. In general, Matching comparisons are more likely to have points above the *kappa* = 80% line than Nonmatching. But in *all* the figures in the Report, Matching or Nonmatching, bullets or cartridges, inconclusives pooled one way or another, the vast majority of the comparisons lie *below* the 80% *kappa* line. In each figure, only a very few examiners rise to having satisfactory consistency as measured by a *kappa* greater than 80%.

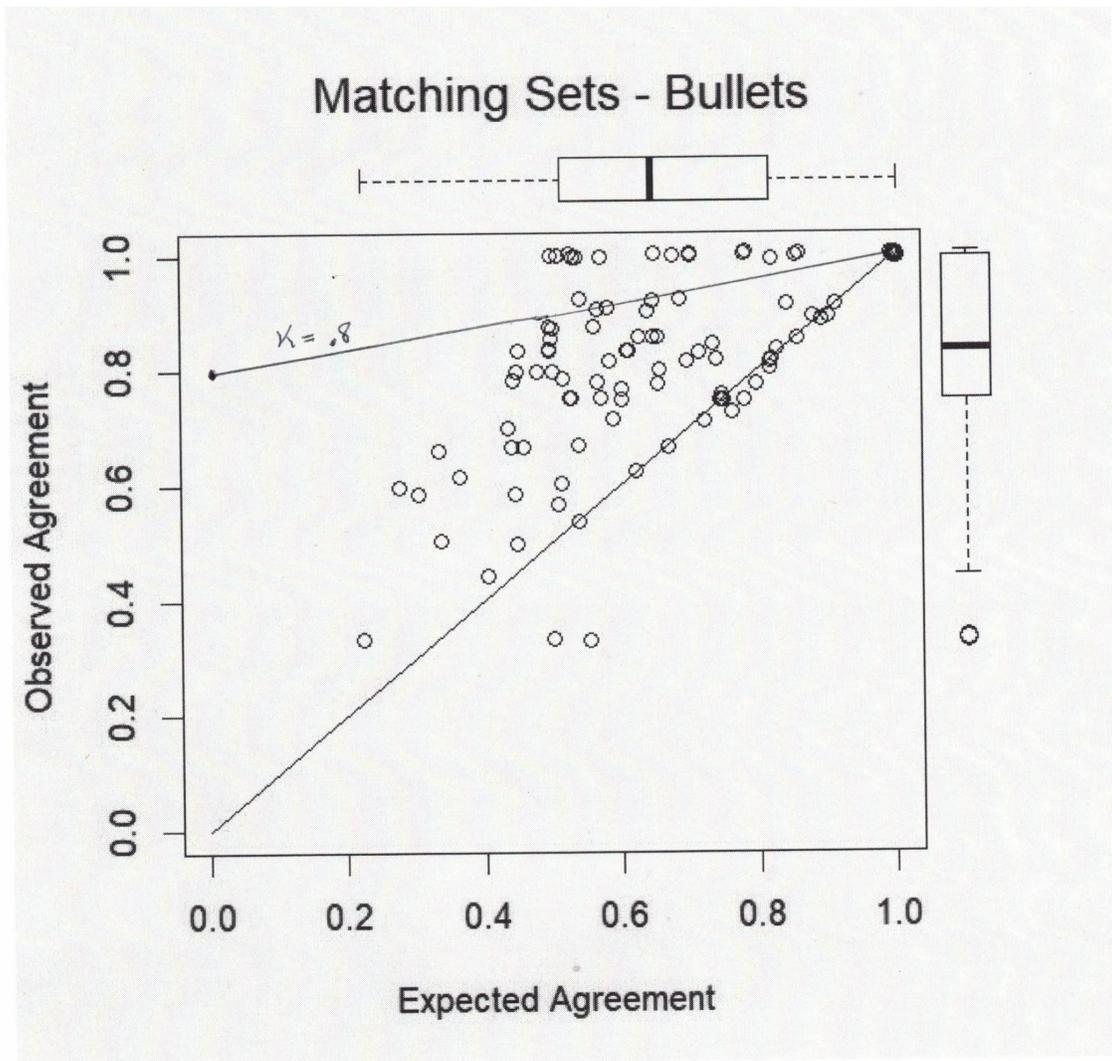

Figure 12a. Repeatability. Inconclusive Categories Pooled

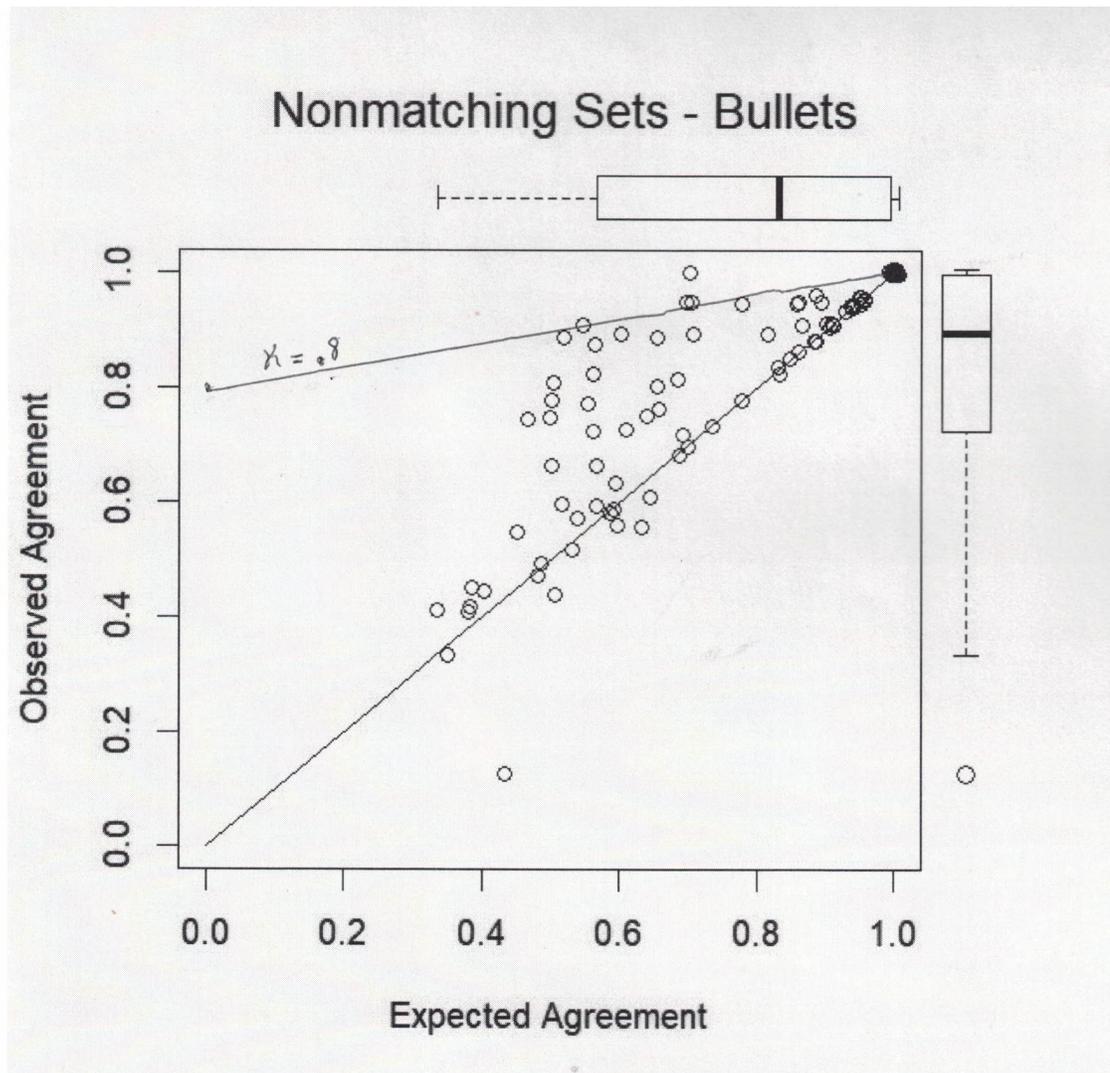

Figure 12b. Repeatability. Inconclusive Categories Pooled

**Conclusion**

The designers of the Ames-USDOE-FBI Study are to be congratulated on an approach that has produced ample data that allow for the assessment of Repeatability and Reproducibility, and for suggesting analytic tools by which to assess the results. It is our conclusion that the tools as they use them are mis-applied and that a better grounded analysis shows, not high Repeatability and Reproducibility as the authors affirm, but, on the contrary, rather weak Repeatability and Reproducibility.

**Appendices**

*A. An example illustrative of expected agreement and Cohen's kappa*

Suppose A, B, and C are three observers of phenomenon P, which takes on, say, three values *b, r, g*, with a regularity known to the observers, say $p_b$ = 10%, $p_r$ = 50%, $p_g$ = 40%, For example, P is a point of light flashing blue, red, green randomly, in the above proportions, known to all three observers. Each of the observers will be exposed to the same series twice, and his/her repeatability assessed.

Suppose A is sharpsighted, and no matter how bright or dim the light, picks up on the color. B is dimsighted, really cannot make out the colors at all and is reduced to making astute guesses in accord with the above proportions. C is somewhere in between: when the light is dim—say about 20% of the time—he is then reduced to guessing and does so mindful of the above proportions.

Each is given a series of 100 flashes of light, followed sometime later by an identical series (long enough after the first that no memory of details survives). The observers don't know more than that the same proportions $p_b$ = 10% etc. hold in the two series.

We look to see how closely each of the three observers comes to repeating themselves on corresponding flashes of light in the two series, i.e. matching first flash to first flash, second to second, etc.. Clearly if the series went something like r, r, g, r, b, g, ....this is exactly what clear-sighted A will report both rounds, and a tabulation of his flash by flash match-up will look like

### Matchups, 1$^{st}$ and 2$^{nd}$ Rounds - A

|  | b | r | g | total first round |
|---|---|---|---|---|
| b | 10 | 0 | 0 | 10 |
| r | 0 | 50 | 0 | 50 |
| g | 0 | 0 | 40 | 40 |
| total second round | 10 | 50 | 40 | 100 |

The overall fraction of times where A agrees with himself—the *observed agreement Po—w*ill be $Po = (10 + 50 + 40)/100 = 100\%$--clearly the maximal amount of self-agreement.

What about B, who can not make out the colors at all and relies on guessing based on the known proportions? Because he is guided by the correct known proportions, he will get the right number of b's etc on both rounds, so the totals on each round should, at least approximately, be in accordance with $p_g$ etc.:

### Matchups, 1$^{st}$ and 2$^{nd}$ Rounds - B

|  | b | r | g | total first round |
|---|---|---|---|---|
| b | ? | ? | ? | 10 |
| r | ? | ? | ? | 50 |
| g | ? | ? | ? | 40 |
| total second round | 10 | 50 | 40 | 100 |

But we can say more: we can expect the fraction of 10 b's on first round that are b, r, and g on the second round to be in accordance with $p_b = 10\%$, etc. So of the 10 first round b's, we can expect 1, 5, and 4 to be b, r, and g on the second round. Thus the above table can be filled in:

## Matchups, 1st and 2nd Rounds - B

|                   | b  | r  | g  | total first round |
|-------------------|----|----|----|-------------------|
| b                 | 1  | 5  | 4  | 10                |
| r                 | 5  | 25 | 20 | 50                |
| g                 | 4  | 20 | 16 | 40                |
| total second round| 10 | 50 | 40 | 100               |

and B's observed agreement is $Po = (1 + 25 + 16)/100 = 42\%$.

What about C? Here we know that the *non*-guesses will be along the diagonal in the 80% of cases where C is perceiving things clearly. Thus at least 8 of the 10 first round b's will also be second round b's, etc:

## Matchups, 1st and 2nd Rounds - C

|                   | b    | r    | g    | total first round |
|-------------------|------|------|------|-------------------|
| b                 | 8+?  | ?    | ?    | 10                |
| r                 | ?    | 40+? | ?    | 50                |
| g                 | ?    | ?    | 32+? | 40                |
| total second round| 10   | 50   | 40   | 100               |

Then, for example, among the 20% of first round b's where guessing takes place—2 instances—we can expect there to be b, r, g in the proportions 1:5:4. About as close that C will come to that is to allot 1 each to r and g. Similarly, for the 20% of reds where guessing takes place—10 instances—we can expect 1, 5, and 4 respectively in the second round, and so on. This yields an overall allotment something like:

### Matchups, 1st and 2nd Rounds - C

|  | b | r | g | total first round |
|---|---|---|---|---|
| b | 8 | 1 | 1 | 10 |
| r | 1 | 45 | 4 | 50 |
| g | 1 | 4 | 35 | 40 |
| total second round | 10 | 50 | 40 | 100 |

and the observed agreement is $Po = (8 + 45 + 35)/100 = 88\%$. Because of reasonable guessing on the 20% unseen, C gets more matches overall than just the 80% where he is seeing clearly. Thus the 12% where C has mismatches underestimates the amount of guessing C is doing.

If we translate the "B Table" into the fraction in each cell, deriving the probability of B's allotting an observation to that cell, we get:

### B's pairings, first to second round fractions

| 1st round | 2nd Round | | | "marginal" 1st round |
|---|---|---|---|---|
|  | b | r | g |  |
| b | 0.01 | 0.05 | 0.04 | 0.1 |
| r | 0.05 | 0.25 | 0.2 | 0.5 |
| g | 0.04 | 0.2 | 0.16 | 0.4 |
| "marginal" 2nd round | 0.1 | 0.5 | 0.4 | 1 |

Because of the random guessing, each entry in the body of the above table is the product of the first round and second round (marginal) fractions. For example, the fraction of time there is a *b* on the first round and *r* on the second is $p_b p_r = 0.1*0.5 = 0.05$ We note that B's observed measure of agreement that we calculated above, $42\% = 0.42$, is just the sum of the fractions on the above diagonal: and can be written $0.1*0.1 + 0.5*0.5 + 0.4*0.4$, the sum of the products of the marginal fractions.

As noted in the main text of this article, for agreement tables like the above, this expression—the sum of the product of corresponding marginal fractions—is called the *expected agreement.*

Since their marginals are the same as B's, the A and C tables have the same expected agreement $Pe = 0.42$. Thus, in particular, to say that C's observed agreement is better than his expected agreement is equivalent to saying his observed agreement is better than what the blind B can achieve.

What about *kappa*?

In the case of A, where $Po = 1$, *kappa*'s upper bound of 1 is achieved. At the other extreme, *kappa* for B is *zero*, since the difference $Po - Pe$ in the numerator of *kappa,* is zero. B achieves nothing beyond what chance can provide.

For C, we have $kappa = (0.88 - 0.42)/(1 - .42) = 0.793 = 79.3\%$, which seems a fair measure of how much non-guesswork C was doing. In fact, we note that $1 - kappa = 20.7\%$ is almost precisely C's percentage of guesswork. This will generally hold under the simple mix of clear-sightedness and guessing we are positing. See Appendix B below.

Needless to say, we do not expect the simple conditions of these examples to hold strictly in practice: judgment comes in all shades of gray between guessing and infallible perception, and guessing is not always in the correct ratios. Nonetheless, our simple model seems to be a helpful way to get a handle on what different values of *kappa* mean.

*B. Cohen's Kappa under perfect global guessing*

To keep notation simple, we consider the situation, as in the example, of three categories, say *b*, *r*, *g*, like for example a light being flashed successively blue or red or green. The argument below will extend to any number of categories. We suppose a sequence of outcomes such as *bbgrggbbb*...of some finite length that gets repeated a second time, either to the same observor after a hiatus long enough for him or her to have forgotten the original, or to a second observor. The instances occur in the proportions $p_b$, $p_r$, $p_g$, where the *p*'s are nonnegative with $p_b + p_r + p_g = 1$, and these proportions are known to the observer(s).

At random moments (same in both series) the observation is obscured (the light is too dim) and the observer makes guesses, keeping the number of guesses of *b* etc. as close as possible to the proportions $p_b$, etc. Guessing happens a fraction $\gamma$ of the time ("gamma" for guessing rate) and the remaining $\pi$ percent of the time the observer judges with pinpoint accuracy ("pi" for precise observation rate). As said, the guesswork occurs at the same moments in the two series. Thus we are assuming there are but two possibilities: pure (but educated) guesswork, or pure perception. We must have $\gamma + \pi = 1$.

Say the series has length *N*. Then in $\pi N$ instances, the observer will observe correctly both rounds and match up in the two series. In the $\gamma N p_b$ instances where the observer guesses *b* in the first round, *b*, *r*, *g* will be guessed on the second round in the proportions $p_b$, $p_r$, $p_g$ respectively. Thus the agreement table will look like

|   | b | r | g |
|---|---|---|---|
| b | $\pi p_b + \gamma p_b^2$ | $\gamma p_b p_r$ | $\gamma p_b p_g$ |
| r | $\gamma p_b p_r$ | $\pi p_r + \gamma p_r^2$ | $\gamma p_r p_g$ |
| g | $\gamma p_b p_g$ | $\gamma p_r p_g$ | $\pi p_g + \gamma p_g^2$ |

For the expected agreement, we need the marginal probabilities.

For first round, marginal for *b* will add up the first row, yielding $\pi p_b + \gamma p_b^2 + \gamma p_b p_r + \gamma p_b p_g =$

$p_b \{\pi + \gamma [p_b + p_r + p_g]\} = p_b$, and likewise for *r* and *g,* and for the columns (2<sup>nd</sup> round marginals)

we also have $p_b, p_r, p_g$ respectively. Thus the expected agreement $P_e = p_b^2 + p_r^2 + p_g^2$.

Summing the diagonal elements of the above table we get the observed agreement

$$P_o = \pi \{p_b + p_r + p_g\} + \gamma \{p_b^2 + p_r^2 + p_g^2\} = \pi + \gamma \{p_b^2 + p_r^2 + p_g^2\} .$$

Thus we get *Cohen's kappa*

$$\kappa = \frac{P_o - P_e}{1 - P_e} = \frac{\pi + (\gamma - 1)\{p_b^2 + p_r^2 + p_g^2\}}{1 - \{p_b^2 + p_r^2 + p_g^2\}} = \frac{\pi \left[1 - \{p_b^2 + p_r^2 + p_g^2\}\right]}{1 - \{p_b^2 + p_r^2 + p_g^2\}} = \pi ,$$

concluding that, in this ideal framework, *kappa* is an exact measure of precise perception/judgment, and 1 – *kappa* is fraction of times the observer is using well tuned guesswork.

**Declaration of Interest**

The first author is a recently retired Senior Mathematical Statistician in the Office of Survey Methods Research at the Bureau of Labor Statistics (BLS) and Senior Consultant at the National Center of Health Statistics. His interest in forensic science occurred as a result of his exposure to a case in which a person who had worked at BLS was convicted of a crime based on firearms comparison evidence. The apparent statistical issues involved spurred his current interest in the subject-matter. The views and conclusions contained herein, however, are those of the authors